\documentclass[%
 reprint,
 amsmath,amssymb,
 aps,
]{revtex4-1}
\usepackage{graphicx}% Include figure files
\usepackage{dcolumn}% Align table columns on decimal point
\usepackage{bm}% bold math
\begin{document}

%\preprint{APS/123-QED}

\title{\boldmath When the Higgs meets the Top: Search for $t \to ch^0$ at the LHC}

\author{Kai-Feng Chen$^{a}$, Wei-Shu Hou$^{a}$, Chung Kao$^{a,b}$, and Masaya Kohda$^{a}$}
 \affiliation{$^{a}$Department of Physics, National Taiwan University, Taipei, Taiwan 10617\\
$^{b}$Homer L. Dodge Department of Physics and Astronomy, University of Oklahoma, Norman, OK 73019, USA
}%Lines break automatically or can be forced with \\


\begin{abstract}
The newly discovered ``Higgs" boson $h^0$, being lighter than the top quark $t$,
%are the two most \emph{massive} and \emph{newest} particles ever.
opens up new probes for flavor and mass generation.
In the general two Higgs doublet model, new $ct$, $cc$ and $tt$ Yukawa couplings
could modify $h^0$ properties.
If $t\to ch^0$ occurs at the percent level,
the observed $ZZ^*$ and $\gamma\gamma$ signal events %for the Higgs boson
may have accompanying $cbW$ activity coming from $t\bar t$ feeddown.
We suggest that $t\to ch^0$ can be searched for via $h^0 \to ZZ^*$, $\gamma\gamma$,
$WW^*$ and $b\bar b$, perhaps even $\tau^+\tau^-$ modes in $t\bar t$ events.
Existing data might be able to reveal some clues for $t \to c h^0$ signature,
or push the branching ratio ${\cal B}(t \to c h^0)$ down to below the percent level.
\begin{description}
% Secondary publications and information retrieval purposes.
\item[PACS numbers]
12.15.Mm, %: Neutral currents
12.60.Fr, %: Extensions of electroweak Higgs sector
14.65.Ha, %: Top quarks
14.80.Ec  %: Other neutral Higgs bosons
%------------------------
% PACS numbers
% You may use the \texttt{description} environment to structure your abstract; use the optional argument of the \verb+\item+ command to give the category of each item.
\end{description}
\end{abstract}

\pacs{Valid PACS appear here}% PACS, the Physics and Astronomy
\maketitle

%\tableofcontents

\section{\label{sec:Motiv}Introduction and Motivation\protect\\}

%\noindent\underline{\it Introduction and Motivation}

With the landmark discovery~\cite{ATL_H, CMS_H} of a 126 GeV boson in 2012
by the ATLAS and CMS experiments at the Large Hadron Collider (LHC),
efforts have shifted towards Higgs property studies, to either
confirm that this is indeed the Higgs boson of the Standard Model (SM),
or to find deviations indicating that it may not be {\it the} Higgs boson.
%
%In this connection, we stress that the Higgs boson and the top quark
%are the two most \emph{massive}, as well as \emph{newest},
%particles ever discovered. %, hence least probed.
With the Higgs boson as the ``mass giver", it is natural to ask
whether it reveals any special effects associated with the
heaviness of the top quark.
A related question is, analogous to the generation repetition
seen with fermions, if we have seen one Higgs boson,
could there be others.

We wish to explore a possible $tch^0$ coupling,
where $h^0$ stands for the 126 GeV boson.
% to indicate that such a coupling is f
Absent within SM,
this coupling has not been the focus of attention.
At the one loop level, the SM branching ratio (for $m_{h^0} = 115$ GeV)
${\cal B}(t \to c h^0) \simeq %4.6 \times 10^{-14}
3\times 10^{-15}$~\cite{Eilam-et-al}
is extremely suppressed.
But with $m_{h^0} < m_t$,
$t\to ch^0$ decay~\cite{Hou91, A-SB00, KCHS12, Craig12}
can readily be searched for in $t\bar t$ production.
In the post-discovery era, % of the 126 GeV boson,
we must search for $t\to ch^0$ at the LHC
as part of the Higgs, and top, property programs.

The $B \to D^{(*)}\tau\nu$ ``anomaly" uncovered by
the BaBar experiment~\cite{BaBarPRL} suggests the need of
a multi-Higgs boson sector.
%Using a ratio method to reduce uncertainties, BaBar
The $B \to D^{(*)}\tau\nu$ rates are found to deviate from SM
expectations, with combined significance over 3$\sigma$.
More intriguing, it cannot~\cite{BaBarPRL} be explained
by the usual two Higgs doublet model (2HDM), type II, the
popular Higgs extension realized in minimal supersymmetry (SUSY).
One explanation is to adopt~\cite{FKNZ12,CGK12} a general 2HDM,
called 2HDM-III, allowing for flavor changing neutral Higgs (FCNH) couplings
%These additional couplings
that modify the charged Higgs couplings,
which could resolve the BaBar anomaly.
The Natural Flavor Conservation (NFC) condition was
proposed over 35 years ago~\cite{NFC} to forbid such couplings,
but the FCNH couplings of Ref.~\cite{CGK12} go even
beyond the Cheng-Sher ansatz~\cite{CS87},
an approach designed to tame FCNH couplings involving lighter quarks.
One basically lets data decide the strength of possible FCNH couplings,
which cannot be argued against.
Note that it was within the Cheng-Sher ansatz that
$t\to ch^0$ (or $h^0 \to t\bar c$) decay was first proposed~\cite{Hou91}
as the leading effect.

\begin{figure}[b!]
\centering
%\vspace{30mm}
{\includegraphics[width=60mm]{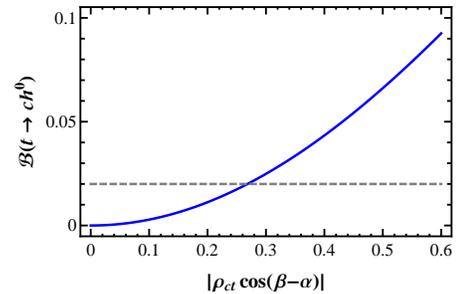}
}
\caption{
${\cal B}(t\to ch^0)$ vs $\rho_{ct} \cos(\beta-\alpha)$,
with 2\% indicated.}
 \label{t-to-ch}
\end{figure}

To account for the BaBar anomaly, the FCNH coupling $\rho_{ct}$
of the exotic heavy Higgs doublet need to be of
order 1 in strength~\cite{CGK12}.
For our purpose, keeping notation of the usual 2HDM-II~\cite{Guide},
the observed boson $h^0$ may contain a small admixture $\cos(\beta-\alpha)$
of the exotic neutral Higgs, hence the $tch^0$ coupling~\cite{hermitian}
\begin{equation}
 \rho_{ct} \cos(\beta-\alpha) \, \bar cth^0 + {\rm h.c.},
\label{eq:tch}
\end{equation}
which can induce $t\to ch^0$ decay. In Fig.~1 we illustrate the branching ratio
${\cal B}(t\to ch^0)$ vs $\rho_{ct} \cos(\beta-\alpha)$.
The question is: Considering all available data, what is the allowed
${\cal B}(t\to ch^0)$, or equivalently, $\rho_{ct}\, \cos(\beta-\alpha)$ value?
What are the signatures to pursue?

We note that, if we take $\rho_{ct} \sim 1$, which is not quite explored
because it is suppressed by $\cos(\beta-\alpha)$ for couplings involving $h^0$
(but not suppressed for couplings involving the heavy exotic Higgs bosons
$H^0$, $A^0$ and $H^{\pm}$), then the analogous parameters
$\rho_{tt}$, $\rho_{cc}$, $\rho_{bb}$ and $\rho_{\tau\tau}$ will
enter the Higgs property study program, as we shall elucidate.
An existing study of multi-lepton final states finds
a bound~\cite{Craig12} of  ${\cal B}(t\to ch^0) < 2.7\%$.
But this should be taken with some caution, as it assumed
SM branching ratios for $h^0 \to WW^*$, $ZZ^*$ and $\tau\tau$ final states.
The study took an effective field theory approach to
isolate the $tch^0$ coupling, hence is not a full theory.

\begin{figure}[t!]
\centering
%\vspace{30mm}
{\includegraphics[width=60mm]{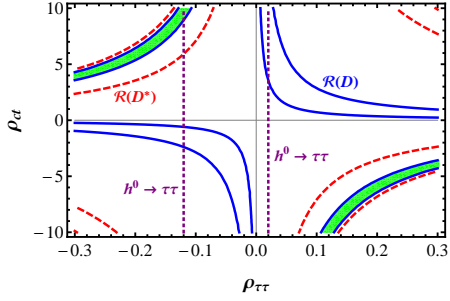}
}
\caption{
Allowed $\rho_{\tau\tau}$--$\rho_{ct}$ region for 2HDM-III
to solve the $B\to D^{(*)}\tau\nu$ anomaly,
taking $m_{H^+} = 700$ GeV (and $\rho_{bb} = 0$).
The shaded-green area is the combined result from
${\cal R}(D)$ (solid-blue lines) and
${\cal R}(D^*)$ (dashed-red lines),
while the dotted-purple lines illustrate
the $h^0\to \tau\tau$ bound by taking
$c_{\beta-\alpha} = 0.2$ in Eq.~(\ref{eq:rho_tautau}).
}
 \label{tautau-ct}
\end{figure}

\section{\boldmath %Implications of $B \to D^{(*)}\tau\nu$,
 BaBar Anomaly, $h^0 \to \tau\tau$, and $b\to s\gamma$}
%\vskip0.3cm
%\noindent\underline{Implications of $B \to D^{(*)}\tau\nu$ Anomaly and $b\to s\gamma$}
%\vskip0.13cm

The BaBar experiment measured the ratios
${\cal R}(D^{(*)})
  = \Gamma(\bar B \to D^{(*)}\tau\nu)/\Gamma(\bar B \to D^{(*)}\ell\nu)$,
finding them both larger than SM expectations,
with a combined significance of $3.4\sigma$.
In 2HDM-II, this implied~\cite{BaBarPRL} that
$\tan\beta/m_{H^+} = 0.44 \pm 0.02$ GeV$^{-1}$ and
$0.75 \pm 0.04$ GeV$^{-1}$ from ${\cal R}(D)$ and  ${\cal R}(D^*)$,
respectively. The impressive precision is because many
uncertainties, both from measurement and from theory, cancel.
The two numbers are incompatible with each other, hence
``excludes the type II 2HDM charged Higgs boson with a 99.8\% confidence level
for any value of $\tan\beta/m_{H^+}$"~\cite{BaBarPRL}.
Either $\tan\beta/m_{H^+}$ value, however, would over-enhance~\cite{FPCP-th}
$B\to\tau\nu$, which is found in agreement (in order of magnitude)
with SM expectation, spelling further trouble.

Employing FCNH parameters in 2HDM-III to account~\cite{FKNZ12,CGK12}
for the BaBar ``anomaly", a new $c$-$t$ coupling
(and $u$-$t$ for $B\to \tau\nu$), heretofore forbidden by
the NFC condition~\cite{NFC}, needs to be of order 1.
For the lepton sector, Ref.~\cite{CGK12} assumed the usual 2HDM-II coupling,
i.e. $\rho_{\tau\tau} = -\tan\beta\sqrt2 m_\tau/v$,
where $v$ is the weak scale.

We shall use the following notation for the Yukawa couplings
of the 2HDM-III Higgs sector~\cite{MS09},
\begin{eqnarray}
&-&{1\over\sqrt2}\sum_{f =}^{ u,\,d,\,\ell}
 \bar f\left[\left(\kappa^f s_{\beta-\alpha}
             + \rho^f c_{\beta-\alpha}\right)h^0 \right. \nonumber\\
&& \quad\quad %\quad\quad\quad\;\;\;
  \left. +\left(\kappa^f c_{\beta-\alpha}
                 - \rho^f s_{\beta-\alpha}\right)H^0
              -i\,{\rm sgn}(Q_f)\rho^f\gamma_5 A^0 \right]f \nonumber\\
&-&\left[\bar u\left(V\rho^dR-\rho^uVL\right)d H^+
  + \bar \nu\rho^\ell R\ell H^+ + {\rm h.c.}\right],
 \label{eq:couplings}
\end{eqnarray}
where $s_{\beta-\alpha}$ ($c_{\beta-\alpha}$) stands for
$\sin(\beta-\alpha)$ ($\cos(\beta-\alpha)$) in
the 2HDM-II notation~\cite{Guide} for sake of comparison
(even though $\beta$ is no longer physical).
The diagonal $\kappa$ terms relate to mass generation,
while the hermitian~\cite{hermitian} $\rho$ terms come from
the second Higgs doublet, %that does not generate any v.e.v.,
and can have off-diagonal terms allowed by data.
It is the combined effect of $\rho_{ct}$ and $\rho_{\tau\tau}$,
both entering through the $H^+$ couplings,
%(in notation of 2HDM II)
that can account~\cite{CGK12} for $B\to D^{(*)}\tau\nu$.
Stringent constraints from down quark sector
imply that only $\rho_{bb}$ needs to be considered;
note that the superscript for $\rho$ can be dropped
for flavor-specific elements.
We shall only keep $\rho_{\tau\tau}$ from lepton sector,
but it is $\rho_{ct}$, $\rho_{tt}$ and $\rho_{cc}$ from
up quark sector that are of interest.

In the so-called ``decoupling limit"~\cite{Guide,Gunion:2002zf},
$\cos(\beta-\alpha) \to 0$ and $\sin(\beta-\alpha) \to 1$,
$h^0$ becomes just the Higgs boson of SM,
while $H^0$, $A^0$ and $H^\pm$ form
an exotic heavy scalar doublet with FCNH couplings.
This limit was tacitly assumed in Ref.~\cite{CGK12},
which advocated $H^0,\;A^0 \to t\bar c$ search.
But we entertain finite $c_{\beta-\alpha}$ values
for sake of $t\to ch^0$ decay, hence would
need to consider $h^0 \to \tau\tau$ data.
The latter from vector boson fusion (VBF) production
is within a factor of 2~\cite{tata_ATLAS} from SM expectations,
\begin{equation}
|s_{\beta-\alpha} + (\rho_{\tau\tau}v/\sqrt2 m_\tau)c_{\beta-\alpha}| \lesssim \sqrt2.
 \label{eq:rho_tautau}
\end{equation}
With $\rho_{\tau\tau} \simeq -0.5$~\cite{CGK12},
even a small $c_{\beta-\alpha}$ could upset this bound.
%
%As we shall see, current LHC data can probe $\rho_{ct}\, c_{\beta-\alpha} \gtrsim 0.2$.
%Keeping $\rho_{ct} \sim 1$, let us insert
To illustrate further, we plot in Fig.~2 the range for
$\rho_{\tau\tau}$--$\rho_{ct}$ allowed by BaBar anomaly
for the typical value of $m_{H^+} = 700$ GeV.
The point $\rho_{\tau\tau} \simeq -0.5$, $\rho_{ct} \sim 1$ of Ref.~\cite{CGK12},
far outside the plot, would require $c_{\beta-\alpha}$ to be rather small.
If we take $c_{\beta-\alpha} = 0.2$ (i.e. $s_{\beta-\alpha} \simeq 0.98$)
in Eq.~(\ref{eq:rho_tautau}), then $-0.12 < \rho_{\tau\tau} < 0.02$
would push $\rho_{ct}$ to become very large, as seen from Fig.~2.

%Smaller $c_{\beta-\alpha}$ would only make the situation worse.

Thus, $h^0\to\tau\tau$ data imply either one
goes to the decoupling limit of $c_{\beta-\alpha} \to 0$,
where $t \to ch^0$ vanishes,
or one has to entertain nonperturbative values for $\rho_{ct}$~\cite{low_mH+}. %, taumu}.
As further analysis~\cite{BaBar13} of $q^2$
($\tau\nu$ pair mass) dependence of $B\to D\tau\nu$
favors New Physics from spin-1 particles,
we will not strongly advocate the link to the BaBar anomaly,
but use it to illustrate that $\rho_{ct}$ can be
of order 1, and focus on probing $tch^0$ coupling directly at the LHC.

\begin{figure}[t!]
\centering
%\vspace{30mm}
{\includegraphics[width=60mm]{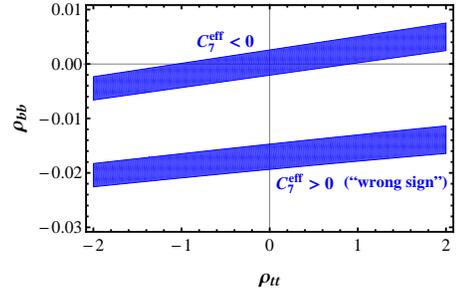}
}
\caption{
Constraint on $\rho_{bb}$ from $b\to s\gamma$,
assuming $\rho_{ct} = 1$, $\rho_{cc} = 0.2$ and $m_{H^+} = 700$ GeV.}
 \label{bsgamma}
\end{figure}

Before turning to the LHC, however, we explore one other piece of
B physics: $b\to s\gamma$ (Fig.~3).
%Although heretofore not well probed,
%One has to be careful with $\rho_{ct} \sim 1$, which is rather large.
We have already seen how $h^0 \to \tau\tau$ data
pushes down $\rho_{\tau\tau}$.
We now show that, if one takes $\rho_{ct} \sim 1$,
the well-known $b\to s\gamma$ process constrains
$\rho_{bb}$ to be rather tiny
(noted recently in Ref.~\cite{CGK13} in a different way).

In the notation of Ref.~\cite{Ciuchini:1997xe}, the $H^+$ loop gives
\begin{align}
\delta C_{7,8}
 \simeq
 \frac{1}{3} & \left( \rho_{tt} +\frac{V_{cs}^*}{V_{ts}^*} \rho_{ct} \right)
   \left( \rho_{tt}^* +\frac{V_{cb}}{V_{tb}}\rho_{ct}^* \right)
  \frac{F_{7,8}^{(1)}(y)}{2m_t^2/v^2} \nonumber \\
 - & \left( \rho_{tt} + \frac{V_{cs}^*}{V_{ts}^*} \rho_{ct} \right) \rho_{bb}
  \frac{F_{7,8}^{(2)}(y)}{2m_tm_b/v^2 },
\end{align}
evaluated at matching scale $\mu_W\sim M_W$, where $y=m_t^2/M_{H^+}^2$
and $F_{7,8}^{(1,2)}(y)$ are given in Ref.~\cite{Ciuchini:1997xe}.
%The point is, $\rho_{bb}$ enters analogous to the $\kappa_{bb}$ (or $m_b$) term,
%hence the loop function is $m_t/m_b$ enhanced.
%It further picks up unsuppressed quark mixing elements
%$V_{cs}^*V_{tb}$ when it is the $\rho_{tc}$ coupling that is
%operative at the $H^+$-$t$-$s$ vertex, which is enhanced
%compared with the similar coefficient of $V_{ts}^*V_{tb}$ in SM.
%
The effect through $\rho_{bb}$ is enhanced by $m_t/m_b$
as well as quark mixing elements,
%Together, these act as lever arms,
such that even a tiny $\rho_{bb}$ could affect $b\to s\gamma$.
We illustrate $\rho_{bb}$ vs $\rho_{tt}$ in Fig.~3,
where we take $\rho_{ct} = +1$, $m_{H^+} = 700$ GeV,
and constrain ${\cal B}(B \to X_s\gamma)$ to be
within 50\% of SM expectation.
The ``wrong-sign" $C_7^{\rm eff}$ case has been included for comparison.
Assuming $C_7^{\rm eff}$ does not change sign,
$|\rho_{bb}|$ is constrained to be considerably less than 0.01.

\section{\boldmath Probing $tch^0$ Coupling at LHC: General 2HDM-III}
%\vskip0.3cm
%\noindent\underline{Probing $tch^0$ Coupling at LHC: General 2HDM-III}
%\vskip0.13cm

With $\rho_{\tau\tau}$ and $\rho_{bb}$ separately
constrained to be small, we are still left with the up-sector
exotic couplings, the FCNH $\rho_{ct}$, as well as the diagonal
$\rho_{cc}$ and $\rho_{tt}$, which we turn to constrain with
present data.
How large can ${\cal B}(t\to ch^0)$ be when $\rho_{ct} \sim 1$?
I.e., what constraint do we have on $\cos(\beta-\alpha)$?
We now take a direct search approach, namely,
from knowledge of top quark physics.

It is clear that ${\cal B}_{ch} \equiv {\cal B}(t\to ch^0)$
cannot be too large, for otherwise we should have seen
deviations in $t\bar t$ measurements during the past two decades.
%, which have very large cross sections.
The best measured $t\bar t$ production cross section
is $\sigma_{t\bar t}^{\ell\ell} \simeq 162$ pb by
CMS~\cite{Chatrchyan:2012bra} at $\sqrt{s} = 7$ TeV
via dileptons, with an experimental error of $5\%$.
On the theoretical side, studies are approaching NNLO
(next-to-next-to-leading order) QCD corrections.
Comparing the two calculations of
Refs. \cite{Moch:2012mk} and \cite{Czakon:2012pz},
theoretical errors up to 10\% appear allowed.
%while ${\cal B}_{ch}$ at 10\% would be unreasonable,
However, a more recent full NNLO result~\cite{Czakon:2013goa}
has reached much better theoretical control.
In any case, given that $\sigma_{t\bar t}^{\ell\ell}$ would be
diluted by $(1-{\cal B}_{ch})^2$,
${\cal B}_{ch}$ of order several percent seems still allowed.

A multi-lepton analysis of Ref.~\cite{Craig12},
based on 7 TeV data from CMS~\cite{Chatrchyan:2012mea},
gives a slightly more stringent bound of 2.7\%, or
$[\sigma\cdot{\cal B}](pp \to t\bar t \to ch^0bW) \lesssim 9$ pb (at 7 TeV).
We have commented that this study assumed SM branching ratios for $h^0$ decay,
which should be taken with caution.
However, it illustrates possible feeddown effects to observable
Higgs boson decay modes, \emph{given the large $t\bar t$ production cross section at the LHC}.
Our chief suggestion is to inspect the clean $ZZ^* \to 4\ell$ samples
of Higgs search data, which we now elaborate.

What has been observed so far at the LHC is
\begin{eqnarray}
\sigma_{gg \to h^0}
 \cdot\frac{\Gamma_{h^0 \to ZZ^*}}{\Gamma_{h^0}^{\rm SM}}
 \cdot\frac{\Gamma_{h^0}^{\rm SM}}{\Gamma_{h^0}}
 \simeq [\sigma\cdot{\cal B}]_{ZZ^*}^{\rm SM},
 \label{eq:4ell}
\end{eqnarray}
where we assume $h^0$ is produced dominantly through gluon-gluon fusion.
We have separated respective pieces where $h^0$-properties
may deviate from SM.
Both experiments find consistency with the expected
15--20 $\ell\ell\ell'\ell'$ signal events expected from
full 2011-2012 data set.
%, with ATLAS (CMS) fluctuating a little higher (lower).
%
However, $\sigma_{t\bar t}$ is of order 220 pb at 8 TeV~\cite{Czakon:2012pz}.
If one takes ${\cal B}_{ch} \simeq 2.7\%$, this amounts
to $\sim 12$ pb into $t\bar t \to ch^0bW$,
which should be compared with~\cite{LHC-H-WG} $\sim 20$ pb
for $gg$-fusion production of a 126 GeV SM Higgs boson!
%One would say there should be
An excess could have appeared already in $ZZ^*$ mode,
except that each of the three product factors in
Eq.~(\ref{eq:4ell}) could deviate from SM.
For example, $\sigma_{gg \to h^0}$ may be smaller,
or $\Gamma_{h^0} > \Gamma_{h^0}^{\rm SM}$ might dilute direct production.

Compared with Ref.~\cite{Craig12}, Eq.~(\ref{eq:couplings})
allows us a more complete treatment of $h^0$-properties
with $\rho_{ct} \sim 1$,
hence understand what SM-like observation of $ZZ^*$ may imply.
Our study also illustrates how 2HDM-III with FCNH
could alter several Higgs properties,
driving in the importance of their measurement.

With $h^0$ dominantly the SM Higgs boson, its $WW^*$ and $ZZ^*$ decay
{rates}, proportional to $\sin^2(\beta-\alpha)$, are hardly changed.
Likewise, the $h^0 \to \gamma\gamma$ rate,
dominated by $W$-loop, is also SM-like.
For fermions, the mass generating $\kappa$ terms are close to SM,
while a small $\cos(\beta-\alpha)$ (we are close to decoupling limit)
dilutes the effect of $\rho$-type couplings.
The consistency of $h^0 \to \tau\tau$ with SM
constrains $\rho_{\tau\tau}$ to be small,
while $\rho_{bb}$ is constrained by $b\to s\gamma$
to be tiny if $\rho_{ct} \sim 1$.
Further diluted by $\cos(\beta-\alpha)$,
the $b\bar b$ rate arises from $\kappa_{bb}$ and is SM-like.

We are left with potential $\rho_{cc}$ and $\rho_{tt}$ effects.
The $c\bar c$ mode is extremely hard to search for,
hence there are no limits so far.
With $\cos(\beta-\alpha) \sim 0.2$,
$\rho_{cc} \sim 0.2$~\cite{rhocc} would bring
$\Gamma_{c\bar c} \sim \Gamma_{b\bar b} \simeq \Gamma_{b\bar b}^{\rm SM}$,
and the enhanced $\Gamma_{h^0}$ would dilute the Higgs signal.
This can be partially compensated for by $\rho_{tt}$,
as this parameter should naturally be of order 1 if $\rho_{ct} \sim 1$,
since $\kappa_{tt} \simeq 1$ also.
With some suppression by $\cos(\beta-\alpha)$,
nevertheless it could bring $\sigma_{gg \to h^0}$ up or down
by a factor of $\sim 2$.

We summarize in Table I possible effects of our
constrained 2HDM-III (with $\rho_{ct} \sim 1$).
While ${\Gamma_{h^0 \to ZZ^*}}/{\Gamma_{h^0}^{\rm SM}}$
is similar to ${\cal B}_{ZZ*}^{\rm SM}$,
$\sigma_{gg \to h^0}$ could change by a factor of 2
and $\Gamma_{h^0}$ could be enhanced.
We comment:
\begin{itemize}
\item If $\rho_{cc}$ is small and $h^0$ branching ratios are SM-like
   (except for $gg$ mode), then the bound of ${\cal B}(t\to ch^0) < 2.7\%$
   from Ref.~\cite{Craig12} would apply;
\item For enhanced $\sigma_{gg \to h^0}$, then dilution of ${\cal B}_{ZZ^*}$
   would be necessary, implying enhanced $h^0\to c\bar c$;
\item If $\sigma_{gg \to h^0}$ is suppressed, or ${\cal B}_{ZZ*}$ is diluted,
   then more $ZZ^*$ events may come from $t\bar t$ feeddown!
\end{itemize}
%
%bound on ${\cal B}(t\to ch)$ would be,
%and it has to be settled by experiment.

\begin{table}[t!]
\label{propHiggs}
\caption[]{Light Higgs $h^0$ properties in 2HDM-III with $\rho_{ct} \sim 1$.
 Widths are in MeV units, with $\Gamma_{h^0}^{\rm SM} \simeq 4.55$ MeV~\cite{HDECAY}.
 %, and Higgs mixing angle $\cos(\beta - \alpha) = 0.2$.
 }
\begin{tabular}{crlcc}
\hline\hline
%&\multicolumn{2}{|c|} {Low Luminosity (10 fb$^{-1}$)}
%& \multicolumn{2}{|c|} {High Luminosity (100 fb$^{-1}$)}  \\
%\cline{1-5}
  & ${\cal B}^{\rm SM}$ \; & \ $\Gamma^{\rm SM}$ & $\Gamma$ & Comment \\
\hline
$WW^*$ & 21.5\% \, & \ 0.98 & \ hard to change\ \ \ & $\sin(\beta - \alpha) \simeq 1$ \\
%\hline
$ZZ^*$ & 2.7\% \, & \ 0.12 & \ hard to change\ \ \ & $\sin(\beta - \alpha) \simeq 1$ \\
%\hline
$\gamma\gamma$ & \quad 0.24\%  & \ 0.011 & \ hard to change\ \ \ & $W$-loop dom. \\
\hline
$bb$ & 59.4\% \, & \ 2.70 & \ hard to change\ \ \ & $b\to s\gamma$ \\
%\hline
$\tau\tau$ & 5.7\% \, & \ 0.26 & within fac. 2 & direct \\
%\hline
$cc$ & 2.6\% \, & \ 0.12 & up to $\sim \Gamma_{b\bar b}$ & not measured \\
&&&& ($\rho_{cc} \lesssim 0.2$) \\
\hline
$gg$ & 7.7\% \, & \ 0.35 & up to fac. 2 & $\rho_{tt} \sim 1$ \\
\hline\hline
\end{tabular}
\end{table}

%An observation can be readily made:
Given the clean signature of the $ZZ^*$ or $\ell\ell\ell'\ell'$ mode,
the searches at CMS and ATLAS have been carried out in an inclusive way,
i.e. simply reconstruct four charged leptons
without looking into any associated byproducts.
The experimental results can thus be used to constrain
any other Higgs production process by looking into
the extra activities in the events,
which already have a clean peak around 126 GeV.
There may well be some fraction of $\ell\ell\ell'\ell' + cbW$ events,
with up to 4 jets.
%extra accompanying $c$-jet plus $b$-jet
%plus ``$W$" activity (be it $\ell\nu$ or dijet).
%\emph{An immediate discovery is possible,
%iff ${\cal B}(t\to ch)$ could be at the several percent level}.
%Conversely, this could quickly set a bound on ${\cal B}(t\to ch)$,
%which is a property of both the top quark and the ``Higgs" boson.
%
%Given the tremendous effort spent on Higgs search,
%this is a much faster track than the more elaborate
%multi-lepton analysis~\cite{Craig12} driven by the $h \to WW^*$ mode,
%which of course can be improved since there is more data.

The CMS preliminary result with full 7 and 8 TeV data~\cite{4l+jets}
shows 13, 8, and 4 events with 0, 1, and 2 jets, respectively,
after selecting events with $m_{4\ell} \in (121.5,\ 130.5)$ GeV.
There is no indication for higher associated jet activity.
To extract a bound on ${\cal B}(t\to ch^0)$, we assume
$\sigma_{gg \to h^0} \cdot {\cal B}(h^0 \to ZZ^*)$ takes SM value.
By inserting the CMS data points, together
with the background histograms provided in the same plot~\cite{4l+jets},
and jet multiplicity distribution from top events,
an upper limit on the top-Higgs contribution is estimated based on
the standard ${\rm CL}_s$ method~\cite{CLs} used at the LHC.
The resulting 95\% confidence level limit on
the relative signal strength between $t \to ch^0$ and inclusive
Higgs production is around 31\%, which can be converted to
a limit of 6.5 pb on the effective cross section of $t \to ch^0$ at 8 TeV,
or a branching ratio limit around 1.5\%.
This result is based on simple jet counting, with no simulation done.

Interestingly, there is in fact one $\ell\ell\ell'\ell' + 4j$ event
observed~\cite{4l+jetsATL} by ATLAS for full 7 and 8 TeV data,
although no jet-multiplicity plot is given.
This event passed the VBF selection, but all 4 jets
(in addition to the 4 leptons) are basically in the central rapidity region.
As an exercise, we simply add 1 more event to the $N_{\rm jet} = 4$ bin,
and without changing anything else,
we obtain an upper limit of 2.2\%, instead of 1.5\%.
If we add 2 events to the $N_{\rm jet} = 4$  bin,
the upper limit becomes 2.8\%.

It is clear that a genuine analysis is best left to the experiments,
as data is already at hand.
The ATLAS event reminds one to carefully check
whether there is any bias towards lower number of jets,
as VBF production is a measurement target.
We remark that, except for our simplifying assumption of
$\sigma_{gg \to h^0} \cdot {\cal B}(h^0 \to ZZ^*)$ being SM-like,
this is in fact a model-independent search for
$t\to ch^0$ ($h^0 \to ZZ^*$) in $t\bar t$ events.

Our argument can be applied to the other mode,
$\gamma\gamma$, that drives the Higgs boson discovery.
But this mode is not so clean, and clearly carries a
bias for VBF event selection of extra jets.
However, for $\gamma\gamma + 4j$ events from $t\bar t$ feeddown,
with $m_{\gamma\gamma}$ in the $m_{h^0}$ window,
the background should be completely different from
the case when jet number is no more than 2,
and should be rather promising.
For the $h^0 \to WW^*$ final state, the multi-lepton analysis
of Ref.~\cite{Craig12} should be redone,
while a specific $\tau\tau + 4j$ analysis can also be pushed.
There is one final ``steadfast" analysis that one could do,
which is searching for $h^0\to b\bar b$ mode
in $t\bar t \to ch^0bW \to cbbb + \ell\nu$.
It has been shown~\cite{KCHS12} that,
through heavy use of $b$-tagging and mass reconstruction,
one should be able to push down to 1\% sensitivity with 2011-2012
data.
Here, ${\cal B}(h^0\to b\bar b)$ might get diluted by $h^0\to c\bar c$,
which was not considered in Ref.~\cite{KCHS12},
but perhaps the actual experimental analysis
could do better than the theoretical study.

It is assuring that, if $h^0$ behaves SM-like
except for inducing $t\to ch^0$ decay, we have
multiple methods to probe ${\cal B}(t\to ch^0)$ down to the 1\% level.
The combined result of the above multi-channel analysis
should reach the sub-percent level,
which becomes comparable with $t\to cZ$ search~\cite{Chatrchyan:2012hqa}.
If the ATLAS $4\ell + 4j$ event is any guide,
we could even make a discovery.

\section{Conclusion}
%\vskip0.3cm
%\noindent\underline{Conclusions}
%\vskip0.13cm

It is of great interest to search for the link between
the top quark $t$ and the Higgs boson $h^0$.
As we have illustrated with $t\bar t \to ch^0bW \to 4l + nj$,
it is quite impressive that the intense efforts of Higgs search
in the past two years could already push the limit on $t\to ch^0$
down to the percent level.
Actual experimental studies of $h^0$ production from $t\bar t$ feeddown,
incorporating
$h^0 \to ZZ^*$, $\gamma\gamma$, $WW^*$, $b\bar b$ and $\tau^+\tau^-$ modes,
should be able to push the limit to below the percent level.
A discovery of the $t\to ch^0$ process with present data
would not only imply the existence of an extended Higgs sector,
but one beyond the usual 2HDM-II of minimal SUSY.

%\

\vskip0.5cm
\noindent{\bf Acknowledgement}.
KFC is supported by the National Science Council grant
NSC 101-2112-M-002-009;
WSH is supported by the Academic Summit grant
NSC 101-2745-M-002-001-ASP,
as well as by grant NTU-EPR-102R8915;
CK is supported in part by the U.S. Department of Energy
under Grant No.~DE-FG02-04ER41305;
MK is supported under NTU-ERP-102R7701 and the Laurel program.
CK thanks the High Energy Physics group of
National Taiwan University for excellent hospitality and
support during a sabbatical visit.
We are grateful to Yang Bai and Bill Murray, as well as others,
for useful discussions.

\vskip0.5cm
\noindent{\bf Note Added}
After submission of this letter, Ref.~\cite{Atwood:2013ica} appeared 
in arXiv which also addresses the $tch^0$ coupling at the LHC.
This study finds a far better sensitivity reach for the $tch^0$ coupling
compared to our results, but we do not understand how it is achieved.
%We do not understand the sensitivity reach of this study.
In addition, after this letter was accepted, we learned
about the ATLAS search for $t\to ch^0$, with $h^0 \to \gamma\gamma$, 
in $t\bar t$ events, finding the limit~\cite{ATLAS-CONF-2013-081} of 
 ${\cal B}(t \to c h^0) < 0.83\%$ at 95\% C.L.


\begin{thebibliography}{9}   % Use for  1-9  references
%\begin{thebibliography}{99} % Use for 10-99 references

%
\bibitem{ATL_H}
  G.~Aad {\it et al.}  [ATLAS Collaboration],
  %``Observation of a new particle in the search for the Standard Model Higgs boson with the ATLAS detector at the LHC,''
  Phys.\ Lett.\ B {\bf 716}, 1 (2012).
%  [arXiv:1207.7214 [hep-ex]].
  %%CITATION = ARXIV:1207.7214;%%

%
\bibitem{CMS_H}
  S.~Chatrchyan {\it et al.}  [CMS Collaboration],
  %``Observation of a new boson at a mass of 125 GeV with the CMS experiment at the LHC,''
  Phys.\ Lett.\ B {\bf 716}, 30 (2012).
%  [arXiv:1207.7235 [hep-ex]].
  %%CITATION = ARXIV:1207.7235;%%

\bibitem{Eilam-et-al}
  G.~Eilam, J.L.~Hewett and A.~Soni,
  %``Rare decays of the top quark in the standard and two Higgs
  %doublet models,''
  Phys.\ Rev.\ D {\bf 44}, 1473 (1991)
  [Erratum-ibid.\ D {\bf 59}, 039901 (1999)];
   %%CITATION = PHRVA,D44,1473;%%
  B.~Mele, S.~Petrarca and A.~Soddu,
  %``A New evaluation of the t ---> cH decay width in the standard
  % model,''
  Phys.\ Lett.\ B {\bf 435}, 401 (1998);
  % [hep-ph/9805498].
  %%CITATION = HEP-PH/9805498;%%
  J.A.~Aguilar-Saavedra,
  %``Top flavor-changing neutral interactions: Theoretical expectations and experimental detection,''
  Acta Phys.\ Polon.\ B {\bf 35}, 2695 (2004).
  % [hep-ph/0409342].
  %%CITATION = HEP-PH/0409342;%%
%
\bibitem{Hou91}
  W.-S.~Hou,
  %``Tree level t ---> c h or h ---> t anti-c decays,''
  Phys.\ Lett.\ B {\bf 296}, 179 (1992).
  %%CITATION = PHLTA,B296,179;%%
  %175 citations counted in INSPIRE as of 14 Mar 2013

\bibitem{A-SB00}
  J.A.~Aguilar-Saavedra and G.C.~Branco,
 %``Probing top flavour-changing neutral scalar couplings
 %  at the CERN LHC,''
  Phys.\ Lett.\ B {\bf 495}, 347 (2000).

%
\bibitem{KCHS12}
  C.~Kao, H.-Y.~Cheng, W.-S.~Hou and J.~Sayre,
  %``Top Decays with Flavor Changing Neutral Higgs Interactions at the LHC,''
  Phys.\ Lett.\ B {\bf 716}, 225 (2012).
%  [arXiv:1112.1707 [hep-ph]].
  %%CITATION = ARXIV:1112.1707;%%
  %2 citations counted in INSPIRE as of 14 Mar 2013

%
\bibitem{Craig12}
  N.~Craig {\it et al.}, %J.~A.~Evans, R.~Gray, M.~Park, S.~Somalwar, S.~Thomas and M.~Walker,
  %``Searching for $t \to c h^0$ with Multi-Leptons,''
  Phys.\ Rev.\ D {\bf 86}, 075002 (2012).
%  [arXiv:1207.6794 [hep-ph]].
  %%CITATION = ARXIV:1207.6794;%%
  %4 citations counted in INSPIRE as of 14 Mar 2013
  We note that $t\to qh$ has recently be discussed by
  R.~Harnik, J.~Kopp and J.~Zupan,
  %``Flavor Violating Higgs Decays,''
  JHEP {\bf 1303}, 026 (2013),
%  [arXiv:1209.1397 [hep-ph]].
  %%CITATION = ARXIV:1209.1397;%%
  with Craig {\it et al.} as the main reference.

%
\bibitem{BaBarPRL}
  J.P.~Lees {\it et al.}  [BaBar Collaboration],
  %``Evidence for an excess of $\bar{B} \to D^{(*)} \tau^-\bar{\nu}_\tau$ decays,''
  Phys.\ Rev.\ Lett.\  {\bf 109}, 101802 (2012).
%  [arXiv:1205.5442 [hep-ex]].
  %%CITATION = ARXIV:1205.5442;%%
  %53 citations counted in INSPIRE as of 14 Mar 2013

%
\bibitem{FKNZ12}
  S.~Fajfer, J.F.~Kamenik, I.~Nisandzic and J.~Zupan,
  %``Implications of Lepton Flavor Universality Violations in B Decays,''
  Phys.\ Rev.\ Lett.\  {\bf 109}, 161801 (2012)
%  [arXiv:1206.1872 [hep-ph]].
  %%CITATION = ARXIV:1206.1872;%%

%
\bibitem{CGK12}
  A.~Crivellin, C.~Greub and A.~Kokulu,
  %``Explaining $B\to D\tau\nu$, $B\to D^*\tau\nu$ and $B\to \tau\nu$ in a 2HDM of type III,''
  Phys.\ Rev.\ D {\bf 86}, 054014 (2012).
%  [arXiv:1206.2634 [hep-ph]].
  %%CITATION = ARXIV:1206.2634;%%
  %26 citations counted in INSPIRE as of 14 Mar 2013

%
\bibitem{NFC}
  S.L.~Glashow and S.~Weinberg,
  %``Natural Conservation Laws for Neutral Currents,''
  Phys.\ Rev.\  D {\bf 15}, 1958 (1977).
  %%CITATION = PHRVA,D15,1958;%%

%
\bibitem{CS87}
  T.P.~Cheng and M.~Sher,
  %``Mass Matrix Ansatz and Flavor Nonconservation in Models with
  %  Multiple Higgs Doublets,''
  Phys.\ Rev.\  D {\bf 35}, 3484 (1987).
  %%CITATION = PHRVA,D35,3484;%%

%
\bibitem{Guide}
  J.F.~Gunion, H.E.~Haber, G.L.~Kane and S.~Dawson,
  %``THE HIGGS HUNTER'S GUIDE,''
  Front.\ Phys.\  {\bf 80}, 1 (2000);
  %%CITATION = FRPHA,80,1;%%
  {\it The Higgs Hunter's Guide} (Addison-Wesley, Redwood City, CA, 1990).

%
\bibitem{hermitian}
  For simplicity and to reduce the number of parameters, we assume
  FCNH couplings to be hermitian.

%
\bibitem{FPCP-th}
  G.W.-S.~Hou,
  Theory Summary talk of ``Flavor Physics and CP Violation" conference,
  Hefei, China, May 2012,
  arXiv:1207.7275 [hep-ph].
  %%CITATION = ARXIV:1207.7275;%%
  %1 citations counted in INSPIRE as of 14 Mar 2013

%
\bibitem{MS09}
  F.~Mahmoudi and O.~St{\aa}l,
  %``Flavor constraints on the two-Higgs-doublet model with general Yukawa couplings,''
  Phys.\ Rev.\ D {\bf 81}, 035016 (2010).
%  [arXiv:0907.1791 [hep-ph]].
  %%CITATION = ARXIV:0907.1791;%%
  %67 citations counted in INSPIRE as of 14 Mar 2013

%\cite{Gunion:2002zf}
\bibitem{Gunion:2002zf}
  J.F.~Gunion and H.E.~Haber,
  %``The CP conserving two Higgs doublet model: The Approach to the decoupling limit,''
  Phys.\ Rev.\ D {\bf 67}, 075019 (2003).
%  [hep-ph/0207010].
  %%CITATION = HEP-PH/0207010;%%
  %184 citations counted in INSPIRE as of 23 Apr 2013

%
\bibitem{tata_ATLAS}
  See, e.g. ATLAS-CONF-2012-160 [ATLAS Collaboration].

%
\bibitem{low_mH+}
  If we adopt smaller $m_{H^+}$, e.g. $\sim 300$ GeV,
  the required $\rho_{ct}$ can be pulled down to perturbative regime.
  With small $m_{H^+}$ and $\rho_{ct}\sim 1$,
  $b\to s\gamma$ data forces $\rho_{bb}$ and $\rho_{tt}$ to be
  significantly smaller than %the corresponding $\kappa$ terms,
  $\kappa_{bb}=\sqrt{2}m_b/v$ and $\kappa_{tt}=\sqrt{2}m_t/v$,
  barring cancellations. We do not pursue such special possibilities
  in this paper.
%
\bibitem{BaBar13}
  J.P.~Lees {\it et al.}  [BaBar Collaboration],
  %``Measurement of an Excess of B -> D(*) Tau Nu Decays and Implications for Charged Higgs Bosons,''
  arXiv:1303.0571 [hep-ex].
  %%CITATION = ARXIV:1303.0571;%%

%
\bibitem{CGK13}
  A.~Crivellin, C.~Greub and A.~Kokulu,
  %``Flavor-phenomenology of two-Higgs-doublet models with generic Yukawa structure,''
  Phys.\ Rev.\ D {\bf 87}, 094031 (2013).
%  arXiv:1303.5877 [hep-ph].
  %%CITATION = ARXIV:1303.5877;%%
  %1 citations counted in INSPIRE as of 07 Apr 2013

%\cite{Ciuchini:1997xe}
\bibitem{Ciuchini:1997xe}
  M.~Ciuchini, G.~Degrassi, P.~Gambino and G.F.~Giudice,
  %``Next-to-leading QCD corrections to B ---> X(s) gamma: Standard model and two Higgs doublet model,''
  Nucl.\ Phys.\ B {\bf 527}, 21 (1998).
%  [hep-ph/9710335].
  %%CITATION = HEP-PH/9710335;%%
  %339 citations counted in INSPIRE as of 06 Mar 2013

%\cite{Chatrchyan:2012bra}
\bibitem{Chatrchyan:2012bra}
  S.~Chatrchyan {\it et al.}  [CMS Collaboration],
  %``Measurement of the $t\bar{t}$ production cross section in the dilepton channel in $pp$ collisions at $\sqrt{s}=7$ TeV,''
  JHEP {\bf 1211}, 067 (2012).
%  [arXiv:1208.2671 [hep-ex]].
  %%CITATION = ARXIV:1208.2671;%%
  %22 citations counted in INSPIRE as of 14 Mar 2013

%\cite{Moch:2012mk}
\bibitem{Moch:2012mk}
  S.~Moch, P.~Uwer and A.~Vogt,
  %``On top-pair hadro-production at next-to-next-to-leading order,''
  Phys.\ Lett.\ B {\bf 714}, 48 (2012).
%  [arXiv:1203.6282 [hep-ph]].
  %%CITATION = ARXIV:1203.6282;%%
  %23 citations counted in INSPIRE as of 14 Mar 2013

%\cite{Czakon:2012pz}
\bibitem{Czakon:2012pz}
  M.~Czakon and A.~Mitov,
  %``NNLO corrections to top pair production at hadron colliders: the quark-gluon reaction,''
  JHEP {\bf 1301}, 080 (2013).
%  [arXiv:1210.6832 [hep-ph]].
  %%CITATION = ARXIV:1210.6832;%%
  %23 citations counted in INSPIRE as of 14 Mar 2013

%
\bibitem{Czakon:2013goa}
  M.~Czakon, P.~Fiedler and A.~Mitov,
  %``The total top quark pair production cross-section at hadron colliders through O(alpha_S^4),''
  arXiv:1303.6254 [hep-ph].
  %%CITATION = ARXIV:1303.6254;%%

%\cite{Chatrchyan:2012mea}
\bibitem{Chatrchyan:2012mea}
  S.~Chatrchyan {\it et al.}  [CMS Collaboration],
  %``Search for anomalous production of multilepton events in $pp$ collisions at $\sqrt{s}=7$ TeV,''
  JHEP {\bf 1206}, 169 (2012).
%  [arXiv:1204.5341 [hep-ex]].
  %%CITATION = ARXIV:1204.5341;%%
  %44 citations counted in INSPIRE as of 14 Mar 2013

%
\bibitem{LHC-H-WG}
  See %LHC Higgs cross section Working Group webpag
  https://twiki.cern.ch/twiki/bin/view/LHCPhysics/\\CERNYellowReportPageAt8TeV.

%
\bibitem{HDECAY}
  Estimates were made using HDECAY,
  A.~Djouadi, J.~Kalinowski and M.~Spira,
  %``HDECAY: A Program for Higgs boson decays in the standard model and its supersymmetric extension,''
  Comput.\ Phys.\ Commun.\  {\bf 108}, 56 (1998),
%  [hep-ph/9704448].
  %%CITATION = HEP-PH/9704448;%%
  which can be contrasted with Ref.~\cite{LHC-H-WG}.

%
\bibitem{rhocc}
  This is not inconsistent with $|\rho_{cc}| \lesssim 0.2$
  found by Ref.~\cite{CGK13} using $D_s \to\tau\nu$, $D_{(s)} \to \mu\nu$
  and $d_n$ (neutron EDM) constraints,
  which suffer from hadronic uncertainties.

%
\bibitem{4l+jets}
  See
  https://twiki.cern.ch/twiki/pub/CMSPublic/Hig130\\02TWiki/jets.pdf
  [CMS Collaboration].

%
\bibitem{CLs}
  A.L. Read, In *Geneva 2000, Confidence limits* 81-101.
  A brief description, especially suitable for theorists,
  can be found in Sec.~II.C of
  J.~Alwall, T.~Enkhbat, W.-S.~Hou and H.~Yokoya,
  %``Doubly resonant WW plus jet signatures at the LHC,''
  Phys.\ Rev.\ D {\bf 86}, 074029 (2012).
%  [arXiv:1208.1686 [hep-ph]].
  %%CITATION = ARXIV:1208.1686;%%

%
\bibitem{4l+jetsATL}
  See Fig. 38 of ATLAS-CONF-2013-013 [ATLAS Collaboration].
  We thank Bill Murray for bringing this to our attention.

%
\bibitem{Chatrchyan:2012hqa}
  S.~Chatrchyan {\it et al.}  [CMS Collaboration],
  %``Search for flavor changing neutral currents in top quark decays in pp collisions at 7 TeV,''
  Phys.\ Lett.\ B {\bf 718}, 1252 (2013).
%  [arXiv:1208.0957 [hep-ex]].
  %%CITATION = ARXIV:1208.0957;%%

%
\bibitem{Atwood:2013ica} 
  D.~Atwood, S.~K.~Gupta and A.~Soni,
  %``Constraining the flavor changing Higgs couplings to the top-quark at the LHC,''
  arXiv:1305.2427 [hep-ph].
  %%CITATION = ARXIV:1305.2427;%%
  %1 citations counted in INSPIRE as of 29 Jul 2013

%
\bibitem{ATLAS-CONF-2013-081}
  ATLAS-CONF-2013-081 [ATLAS Collaboration].


\end{thebibliography}
\end{document}